\documentclass{article}
\usepackage{dcase2021,amsmath,graphicx,url,times,booktabs, tabularx}
\usepackage{enumitem}
\usepackage{float}
\usepackage{multirow}
\usepackage{booktabs}
\usepackage{xcolor}
\usepackage{color, colortbl}

\definecolor{Gray}{gray}{0.9}

\title{Squeeze-Excitation Convolutional Recurrent Neural Networks for Audio-Visual Scene Classification}

\name{Javier Naranjo-Alcazar$^{1,2}$,
      Sergi Perez-Castanos$^{2}$,
      Aaron Lopez-Garcia$^{2}$,
      Pedro Zuccarello$^{1}$,
      }
\secondlinename{	  
      Maximo Cobos$^{2}$,
      Francesc J. Ferri$^{2}$
      }

\address{$^1$ Instituto Tecnológico de Informática, València, Spain \{jnaranjo, pzuccarello\}@iti.es\\         
$^2$ Universitat de Val\`encia, Burjassot, Spain, \{pecaser@alumni.uv.es, \{maximo.cobos, francesc.ferri\}@uv.es\}\\
 }

\begin{document}

\ninept
\maketitle

\begin{sloppy}

\begin{abstract}
The use of multiple and semantically correlated sources can provide complementary information to each other that may not be evident when working with individual modalities on their own. In this context, multi-modal models can help producing more accurate and robust predictions in machine learning tasks where audio-visual data is available. This paper presents a multi-modal model for automatic scene classification that exploits simultaneously auditory and visual information. The proposed approach makes use of two separate networks which are respectively trained in isolation on audio and visual data, so that each network specializes in a given modality. The visual subnetwork is a pre-trained VGG16 model followed by a bidiretional recurrent layer, while the residual audio subnetwork is based on stacked squeeze-excitation convolutional blocks trained from scratch. After training each subnetwork, the fusion of information from the audio and visual streams is performed at two different stages. The early fusion stage combines features resulting from the last convolutional block of the respective subnetworks at different time steps to feed a bidirectional recurrent structure. The late fusion stage combines the output of the early fusion stage with the independent predictions provided by the two subnetworks, resulting in the final prediction. We evaluate the method using the recently published TAU Audio-Visual Urban Scenes 2021, which contains synchronized audio and video recordings from 12 European cities in 10 different scene classes. The proposed model has been shown to provide an excellent trade-off between prediction performance (86.5\%) and system complexity (15M parameters) in the evaluation results of the DCASE 2021 Challenge.

% Automatic scene classification has always been one of the core tasks in every edition of the DCASE challenge. Until this edition, such classification was performed using only audio data, and so the problematic was defined as Acoustic Scene Classification (ASC). In this 2021 edition, audio data is accompanied with visual data, providing additional information that can be jointly exploited for achieving higher recognition accuracy. The proposed approach makes use of two separate networks which are respectively trained in isolation on audio and visual data, so that each network specializes in a given modality. After training each network, the fusion of information from the audio and visual subnetworks is performed at two different stages. The early fusion stage combines features resulting from the last convolutional block of the respective subnetworks at different time steps to feed a bidirectional recurrent structure. The late fusion stage combines the output of the early fusion stage with the independent predictions provided by the two subnetworks, resulting in the final prediction. For the visual subnetwork, a VGG16 architecture pretrained on the Places365 dataset is used, applying a fine-tuning strategy over the Challenge dataset. On the other hand, the audio subnetwork is trained from scratch and uses squeeze-excitation techniques as in previous contributions from this team.  As a result, the final accuracy of the system is 92\% on development split, outperforming the baseline by 15 percentage points.

\end{abstract}


\section{Introduction}
\label{sec:intro}

\par The world as it is perceived by humans involves multiple modalities. In general, a sensory modality is understood as a primary channel of communication and sensation, such as vision, hearing or touch. In this context, multi-modal machine learning aims at exploiting datasets including multiple such modalities, building models that can process and relate information among them \cite{baltruvsaitis2018multimodal}. The explosion of deep learning and its use in vision, natural language processing and acoustic analysis, makes of multi-modal machine learning a multi-disciplinary area with increasing potential. Obviously, the most abundant multi-modal datasets are those made up of audio-visual data, where the included examples come in the form of videos that include both sound and images. 

One important application scenario of multi-modal machine learning is that of automatic scene classification on videos, which refers to the task of classifying audiovisual data to one of the predefined scene categories (such as airport, park or shopping mall) \cite{zeng2021deep}, based on the ambient content provided by the information contributed by both modalities. Note, however, that scene classification has also been a topic of intensive research in the last decade by considering different modalitites on their own \cite{lowry2015visual,abesser2020review,cheng2017remote}.

\par This paper presents a multi-modal approach for scene classification consisting of two specialized components or modules (an audio module and a visual module) that are further trained together to achieve a more robust solution by incorporating  two fusion strategies simultaneously. The visual module is based on a VGG16 \cite{simonyan2014very} convolutional neural network (CNN) pre-trained on the \emph{places365} dataset \cite{zhou2017places, gkallia2017keras_places365}. The training procedure of this component is based on a transfer learning scheme with fine tuning. On the other hand, the audio module is based on a fully convolutional neural network with convolutional blocks implementing residual and squeeze-excitation techniques \cite{naranjo2020acoustic} and gammatone filterbank audio representations as input. Finally, the audio and video modules with frozen weights are combined into a multimodal recurrent structure that performs information fusion both at early and late stages. The early fusion stage combines features resulting from the last convolutional block of the audio and visual subnetworks at different time steps, while the late fusion stage provides a final prediction by combining the output of the early fusion stage with the independent predictions provided by the visual and audio modules.

The model is evaluated by considering the TAU Audio-Visual Urban Scenes 2021 dataset \cite{Wang2021_ICASSP}, which contains synchronized audio and video recordings from 12 European cities in 10 different scenes classes. For a complete assessment of the model, different input pre-processing alternatives and architecture choices are considered and discussed. 

The rest of this paper is structured as follows. Section~\ref{sec:method} describes in detail the architecture of the different sections making up the whole learning system. Section~\ref{sec:experiments} describes the experimental set-up and evaluates the proposed system considering some variants with respect to the input and the system architecture. Finally, Section~\ref{sec:conclusion} concludes this work.

\begin{figure*}[t]
    \centering
    \centerline{\includegraphics[width=\textwidth]{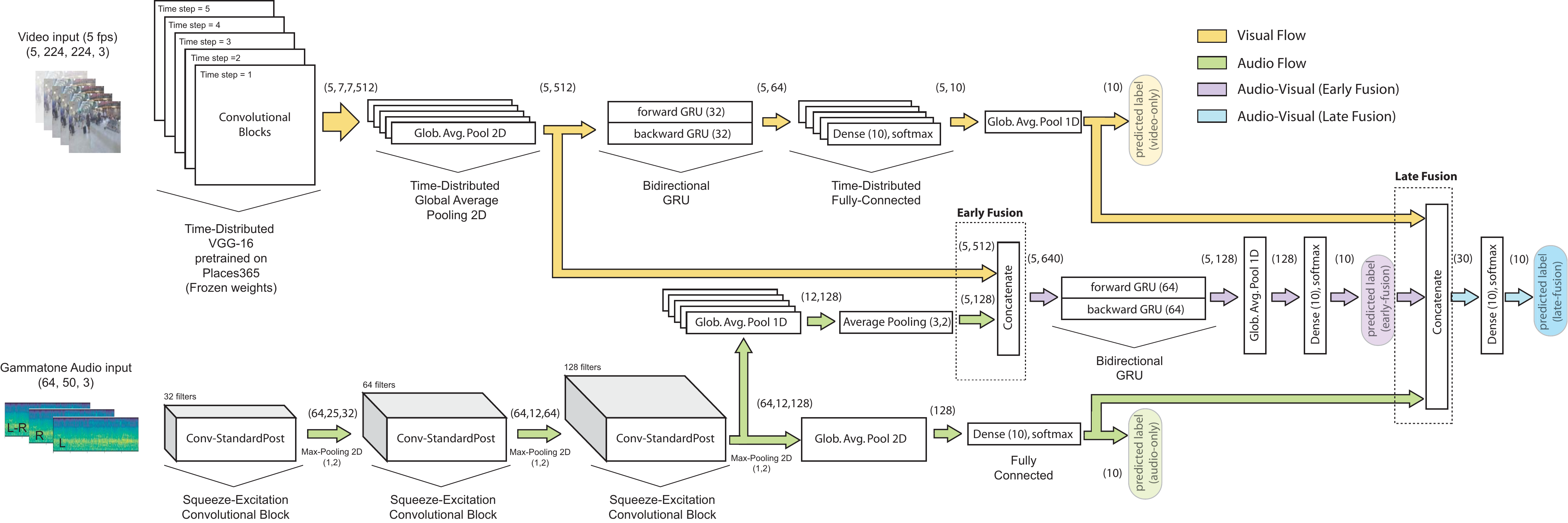}}
    \caption{Proposed network achitecture for audiovisual scene classification.}
    \label{fig:system}
\end{figure*}

\section{System Description}\label{sec:method}

This section describes the full architecture of the system, providing details on the different modules making up the whole multi-modal network. An schematic view of the full model is depicted in Fig.~\ref{fig:system}, where the visual and audio flows are represented with different colors.

\subsection{Audio Module}\label{subsec:audio_repr}

\subsubsection{Audio Input Representation}

\par Based on previous works by the authors \cite{perezcastanos2019cnn, naranjo2020task}, the input to the network consists in a multi-channel 3D audio representation compiling information from the left and right input audio channels as well as their difference. Each channel is converted into a time-frequency domain representation, provided by a Gammatone or a Mel-scale filter bank. Both alternatives have been widely adopted by the machine listening community \cite{tabibi2017investigating, zhang2018deep, Perez-Castanos2020a, Perez-Castanos2020b} and we evaluate both options in the experimental section of this paper.

\par In any case, all the considered representations are computed using 64 frequency bands, with a window size of 40 ms and 50\% overlap. The audio was resampled from 48 kHz to 44.1 kHz. Gammatone representations were computed by using the Auditory Toolbox presented in \cite{slaney1998auditory} with Python implementation, while Mel-spectrograms were obtained by using the LibRosa library \cite{mcfee2015librosa}. Taking the above details into account, one second of audio results in a tensor input of size (64, 50, 3), where the third axis corresponds to the left-right-difference channels.

% one spectrogram is obtained for each audio clip. This spectrogram is then divided into 10 spectrograms corresponding to 10 spectrograms of one second each.

\subsubsection{Audio Subnetwork}\label{subsec:SE}

\par The audio module is based on a fully convolutional neural network combining residual connections with squeeze-excitation. More specifically, the convolutional blocks follow the structure of those denoted as \emph{Conv-StandardPOST} in \cite{naranjo2020acoustic} (see~Fig.~\ref{fig:squeeze}), which showed very good performance for acoustic scene classification tasks. The aim of these blocks is to achieve improved accuracy by recalibrating the internal feature maps using residual \cite{he2016deep} and squeeze-excitation techniques \cite{hu2018squeeze, roy2018concurrent}. An important feature is the use of the scSE (spatial and channel Squeeze and Excitation) module, which performs a spatial and channel-wise recalibration of the block feature maps. The interested reader is referred to \cite{naranjo2020acoustic} for a full description and evaluation of such blocks. All use a $3\times 3$ kernel size, while the number of filters in each block are specified in Fig.~\ref{fig:system}. In between convolutional blocks, Max Pooling layers are used to halve the resolution of the resulting feature maps along the time axis. Additionally, Dropout \cite{srivastava2014dropout} with a rate of $0.3$ is also included after the pooling layers to prevent overfitting. The output feature maps from the last convolutional block are summarized with global average pooling into a 128-dimensional feature vector, which is fed to a fully-connected layer with softmax activation for classification. This subnetwork is trained from scratch using only audio data on the whole dataset by minimizing the cross-entropy loss.

\begin{figure}[t]
    \centering
    \centerline{\includegraphics[width=0.4\columnwidth, angle =90]{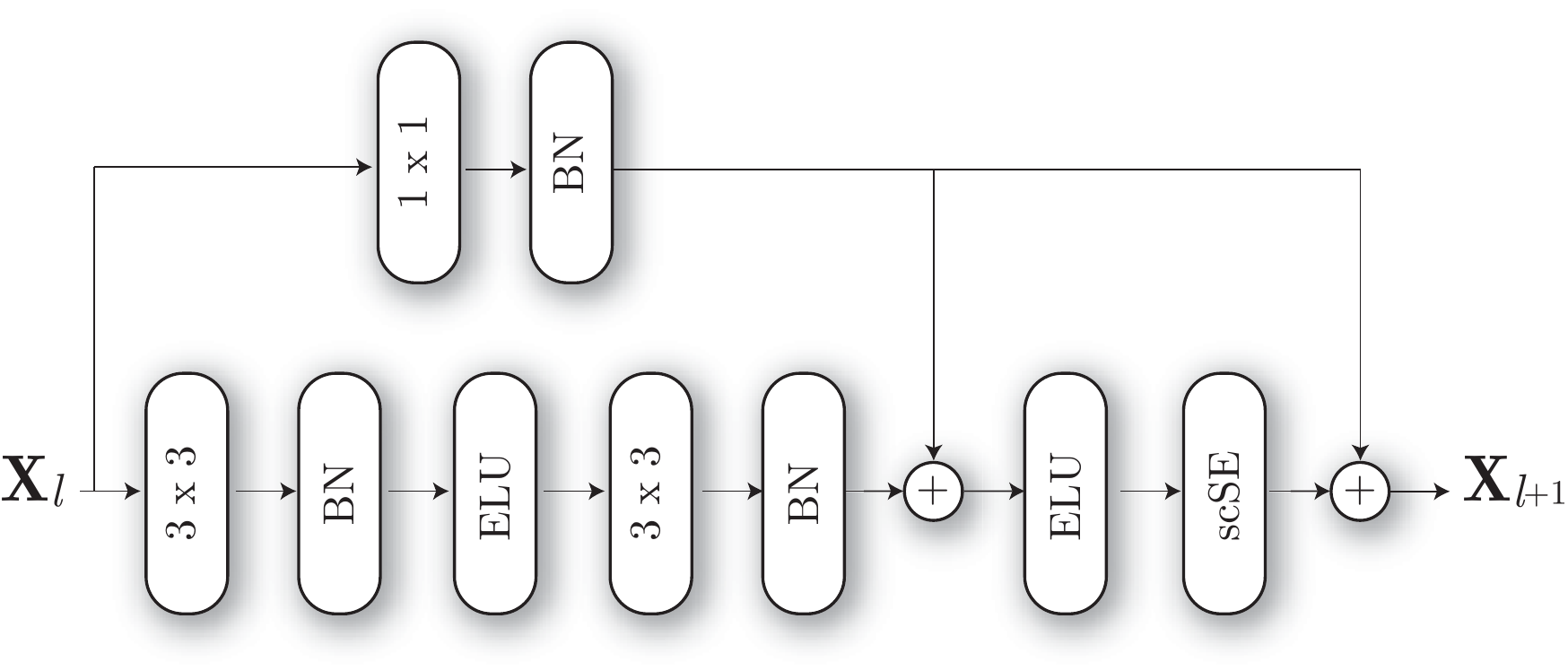}}
    \caption{Structure of the \emph{Conv-StandardPOST} block \cite{naranjo2020acoustic}. BN and ELU denote batch normalization and exponential linear unit activation, respectively. The $N \times N$ notation denotes a convolutional layer with the corresponding kernel size. The input to the $l$-th block is denoted as $\mathbf{X}_l$.}
    \label{fig:squeeze}
\end{figure}

%\par Most Acoustic Scene Classification framework rely on the ability of a CNN to extract meaningful features. Either in a VGG-style \cite{simonyan2014very} or Residual \cite{he2016deep} networks are very similar between different submissions. Therefore, the improvement of the systems often falls on other aspects such as data augmentation techniques (mixup \cite{zhang2017mixup} and temporal cropping among others) or the ensemble of many independent models \cite{han2017convolutional, chen2019integrating, koutini2019cp}. 

%\par In \cite{naranjo2020acoustic} an analysis of different residual-excitation blocks proposed in \cite{hu2018squeeze} plus the contribution of two novel blocks using the \textit{Concurrent Spatial and Channel Squeeze and Channel Excitation} configuration presented in \cite{roy2018concurrent} is carried out in ASC task. The analysis is run without any data augmentation technique. According to \cite{naranjo2020acoustic}, the novel \textit{Conv-StandardPOST} configuration shows the best results in ASC problems. All networks used for this submission incorporate this residual-excitation block. For more insight of this choice, see \cite{naranjo2020acoustic}.

\subsection{Visual Module}\label{subsec:visual}

\subsubsection{Visual Input Representation}

The visual input is adapted to match the pre-trained VGG16 architecture \cite{simonyan2014very}, which accepts color images of size 224$\times$224 pixels. Moreover, as visual scene recognition does not require a very high frame rate (images do not change that much from frame to frame), the videos from the dataset are subsampled for obtaining a frame rate of 5 frames per second (fps). Therefore, a one-second video clip results in a tensor shape of (5, 224, 224, 3).

\subsubsection{Visual Subnetwork}

\par The visual module is based on the VGG16 CNN architecture \cite{simonyan2014very} pretrained on the \emph{places365} dataset \cite{zhou2017places}. With the aim of processing temporal information extracted from multiple frames, a time-distributed structure with frozen weights is considered. It must be emphasized that the pre-trained model is used as a feature extractor, so that the top fully-connected layers of the network are omitted. The outputs from each time step (5 temporal steps) are globally averaged channel-wise, resulting in a sequence of 512 output features. This sequence is fed to a bidirectional 64-neuron Gated Recurrent Unit (GRU) layer and the returned sequences are processed by a time-distributed fully-connected layer with softmax activation, resulting in a predicted label for each time step. The final label is taken as the temporal average of the predictions. This subnetwork is trained on the visual data only, with trainable weights only on the recurrent and final dense layer.

% Regarding the visual module, it has been decided to approach the training with a strategy known as transfer learning. This technique is based on prior knowledge of the neural network. This prior knowledge is obtained by training the neural network with another/external dataset. The training of this network is used as a starting point for training the system to the Challenge dataset.

% \par The dataset used is known as places365 \cite{zhou2017places}.  This dataset is composed of images from several scenes, 365 classes. The implementation we have used as a starting point can be found at the following link\footnote{https://github.com/GKalliatakis/Keras-VGG16-places365}. In such an implementation, a VGG network \cite{simonyan2014very}, namely VGG16, is trained with such a dataset.

% \par This implementation corresponds to the first block of the visual module. As can be seen in Figure~\ref{fig:asc}, the weights of the network are frozen, they are not modified by the Challenge dataset information. It should be noted that the convolutional layers have been modified to TimeDistributed in order to process temporal information.

% \par It has been decided to feed the module with 5 images per second. In order to obtain temporal relationships between the 5 frames, a GRU recurrent layer has been stacked. Thus, the network corresponds to a CRNN architecture. After the recurrent layer, a Dense TimeDistributed layer is added, which is used for the final classification, together with a Global Average Pooling layer and the final classification layer.

\subsection{Full Audio-Visual Network}

The complete audio and visual modules described above are then merged into a full audio-visual framework that combines information from both modalities at two different levels. On an early fusion stage, the output of the last convolutional block of the audio and visual modules are concatenated into a sequence of 640 features. To achieve this, the feature maps of the audio module are turned into a temporal sequence matching the temporal resolution of the visual data (i.e. 5 fps) using global and average pooling operators. A bidirectional GRU processes the sequence, and a new prediction is created by stacking a global average pooling and a dense layer. A late fusion stage receives the predictions from the independent modalities as well as the one resulting from their combination and produces the final prediction with a dense layer with softmax activation.

Note that, as observed in Fig.~\ref{fig:system}, the full network can be used to extract both predictions from the independent modalitites, i.e. labels from the visual (yellow) and audio (green) information flows, and from the fusion flows, i.e. early fusion (purple) and late fusion (blue). 

\subsection{Dataset}

The system is trained on the recently published TAU Urban Audio-Visual Scenes 2021 \cite{}. This dataset contains fragments of recordings obtained in 12 large European cities corresponding to 10 scene classes: airport, shopping mall (indoor), metro station (underground), pedestrian street, public square, street (traffic), traveling by tram, bus and metro (underground), and urban park. The data was gathered with four devices recording simultaneously. The data examples are provided as segments with a length of 10 seconds, annotated by the corresponding scene class, city and recording location identifier. The dataset contains 34 hours of recordings and it is accompanied by training/test setups to facilitate comparisons. The training set contains approximately 70\% of the data, while the validation set contains the remaining 30\%.

\subsection{Training Details}\label{subsec:training}

The whole network was trained in three steps using the default training and validation partitions provided by the TAU Audio-Visual Urban Scenes 2021 dataset. The first step corresponds to the training of the audio module from scratch using audio data only. The second step trains the recurrent and classification parts of the visual module (the convolutional blocks use frozen weights from the pre-trained network). In the last step, the whole audio-visual network is trained using frozen weights from the audio and visual modules. A fine-tuning strategy is finally followed, unfreezing all the weights and using a very small learning rate. The loss function used at each training step was categorical cross-entropy. The optimizer used was Adam \cite{kingma2014adam} with default parameters. The models were trained with a maximum of 200 epochs. Batch size was set to 32 for training the independent subnetworks and 16 for the complete audio-visual network due to memory constraints. The 10 second examples provided in the dataset were randomly trimmed into 1 second segments in each epoch. The learning rate started with a maximum value of 0.001 decreasing with a factor of 0.5 in case of no improvement in validation accuracy after 20 epochs. In the last fine-tuning with all trainable weights, the starting learning rate was $10^{-5}$. The training is considered as early finished in case of no improvement in validation accuracy after 50 epochs. Mixup data-augmentation \cite{zhang2017mixup} with $\alpha = 0.4$ was used. All the models were implemented using Keras with Tensorflow backend and trained using NVidia Titan RTX GPU.

\begin{table*}[]
\centering
\begin{tabular}{ccccc}
\hline
\rowcolor{Gray} \multicolumn{1}{l}{} & \multicolumn{4}{c}{Modality} \\ \hline
 & \textbf{Audio-Only} & \textbf{Visual-Only} & \textbf{Multi-Modal (Early Fusion)} & \textbf{Multi-Modal (Late Fusion)} \\
\emph{log-Mel} & 68.4 & 87.0 & 88.5 &  88.7 \\
\emph{Gammatone} & 69.0 & 87.0 & 89.2 & 90.0 \\ \hline
\end{tabular}
\caption{Accuracy results on the TAU Audio-Visual Urban Scenes 2021 validation partition.}
\label{tab:dev_results}
\end{table*}

\begin{table*}[]
\centering
\begin{tabular}{ccc}
\hline
\rowcolor{Gray} \multicolumn{3}{c}{Modality} \\ \hline
\textbf{Audio-Only (Gammatone)} & \textbf{Visual-Only} & \textbf{Multi-Modal (Late Fusion)} \\
 66.8 & 83.2 & 86.5 \\ \hline
\end{tabular}
\caption{Accuracy results on the DCASE 2021 Task1b evaluation set.}
\label{tab:eval_results}
\end{table*}

\subsection{Model Complexity}\label{subsec:complexity}

Assuming that all the weights of the different subsystems are trainable, the number of parameters corresponding to the different modules of the network are as follows: audio module (323k trainable weights), visual module (14M parameters, only 105k trainable) and full audio-visual system (15M parameters, only 272k trainable). Note that, although the number of parameters used by the visual module is considerably higher than that of the audio module, the visual one uses frozen weights in all its convolutional blocks. Thus, all subsystems can be trained considerably fast, as only 272k weights are trainable. Additionally, the final fine-tuning step where all the weights are unfrozen only requires 2 epochs and, therefore, it does not require too much extra training time.

\section{Experiments}\label{sec:experiments}

This section evaluates the proposed multi-modal framework over the default validation partition of the TAU Audio-Visual Urban Scenes 2021 dataset. For those systems submitted to the DCASE 2021 Challenge, we also provide the performance reported by the organizers over the evaluation/test dataset. Note that no custom test partition was created in order to facilitate comparisons with other competing systems using the same dataset. The performance of the proposed system is analyzed considering three different aspects:
\begin{itemize}
    \item Audio input representation: log-Mel spectrogram and gammatone filterbank.
    \item Independent modalities: audio-only and visual-only.
    \item Multi-modal fusion: early fusion and late fusion.
\end{itemize}

Table~\ref{tab:dev_results} shows the accuracy results obtained for the different subsystems involved in the audio-visual framework on the default validation set. Similarly, Table~\ref{tab:eval_results} shows the results obtained for the evaluation partition used in DCASE 2021 Task1b. Comparisons to other competing approaches can be directly accessed via the DCASE 2021 Challenge website\footnote{\url{http://dcase.community/challenge2021/task-acoustic-scene-classification-results-b}}.

In general, the results clearly highlight that the visual-only modality is much more accurate than the audio-only one (e.g. 87.0\% vs 69.0\% in the development validation partition). Although it is true that the visual network departs from previous knowledge provided by a pre-trained model, this confirms that, as of today, acoustic scene recognition is a more challenging problem than visual scene recognition. Nonetheless, the audio module used in this work was ranked as the best performing model from all the submissions considering only the audio modality in terms of log-loss performance. Also, it is observed that although both audio input representations, log-Mel and Gammatone spectrograms, led to very similar performances, the results were slightly better with the use of Gammatone filterbanks (69.0\% vs 68.4\% in the validation partition).

Another interesting observation is that the multi-modality performance, despite being always better than that of the individual modalities alone, seems to be only slightly better than the best of the considered modalities, in this case, the visual one. In any case, for our specific case and despite the significant performance gap between the audio and video modalities, the multi-modal approaches achieved a performance gain of approximately 3 percentage points. In this context, although both the early and late fusion stages were able to exploit the information from the audio and visual data, the late fusion stage performed consistently (slightly) better in all our experiments.

To provide further insight, Fig.~\ref{fig:class-wise} shows the performance achieved by each modality and their combination in the DCASE 2021 Task1b evaluation set on each class. Note that the audio module only outperforms the visual one for the tram class and it can be clearly observed for this case that multi-modality allows to exploit significantly both types of information. Interestingly, although the multi-modal system usually outperforms the individual modalities alone, there are some classes at which this does not happen, as in public square or metro.

\begin{figure}[t]
    \centering
    \centerline{\includegraphics[width=\columnwidth]{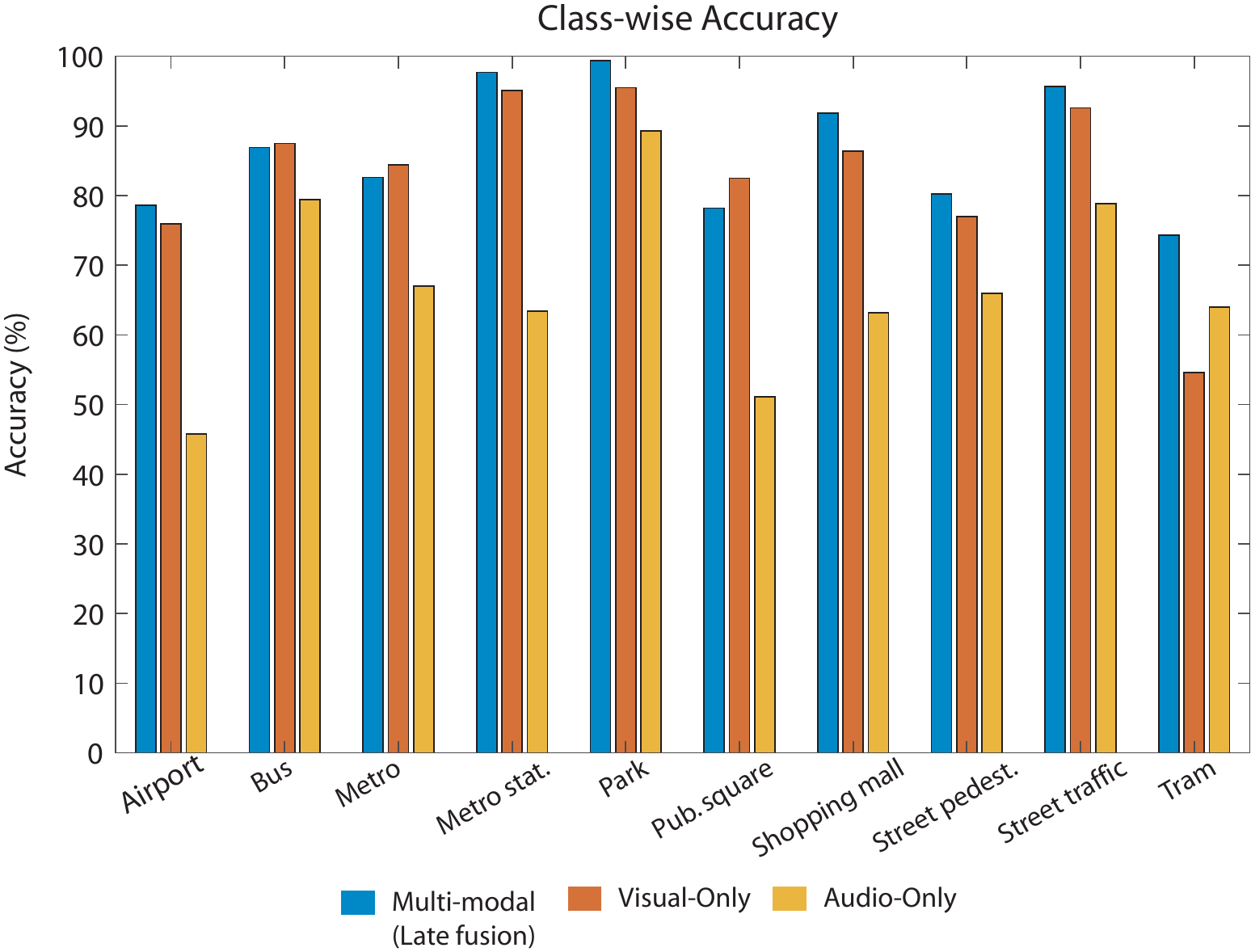}}
    \caption{Class-wise performance in DCASE 2021 Task1b evaluation set.}
    \label{fig:class-wise}
\end{figure}

% \section{Results}\label{sec:results}

% \par The results obtained by the scene classification system, using each module separately and together, are shown below.
% As observed in Table~\ref{tab:results}, the baseline is exceeded in all 3 cases. The audio network improves the baseline by 4 percentage points while the visual network improves the baseline by about 22 points. As can be noticed (and this being the aim of this Challenge) merging both sources of information leads to a more accurate system. The best performing module (visual) improves by almost 4 points when combined with auditory information, leading to a final accuracy of 90.0\%.

% \begin{table}[H]
% \centering
% \begin{tabular}{cccc }
% \toprule
%  & Audio & Visual & Audio-Visual  \\ 
%  \midrule
% \rowcolor{Gray} Challenge Baseline & 65.1 & 64.9 & 77.0 \\ 
%  \midrule
%  Proposed system & 69.0 & 86.5 & 90.0 \\
%  \bottomrule
% \end{tabular}
% \caption{Accuracy (\%) results obtained compare with the proposed baseline}
% \label{tab:results}
% \end{table}

\section{Conclusion}\label{sec:conclusion}

This paper presented a multi-modal system for audio-visual scene classification based on convolutional recurrent neural networks. The full system is based on two individual modules that are trained in isolation on the audio and visual modalities, respectively. The visual module is based on a time-distributed VGG16 model pre-trained on the places365 dataset, followed by a bidirectional recurrent layer. The audio module is a convolutional neural network that incorportes residual and squeeze-excitation techniques, working over Gammatone input representations. After training both modules, both are incorporated into a full audio-visual architecture that performs information fusion at early and late stages. Early fusion combines features extracted from both modalities at each time step into a bidirectional recurrent layer. Late fusion decides a final label after receiving the predictions obtained from each independent modality and that resulting from early fusion. The results show that the proposed framework is able to exploit successfully information from both modalities, even though the visual modality is considerably more accurate than the audio one. The results obtained in the DCASE 2021 Challenge confirm that the proposed system provides an excellent trade-off between prediction performance and system complexity.

% \par Nowadays, there are a multitude of machine learning solutions adapted to data from different domains (image, speech, audio, ...). However, mixed solutions exploiting jointly data from multiple domains are not as well explored. Task 1b of the 2021 edition of the DCASE Challenge proposes a modification of the classic ASC task to turn it into an audio-visual task where, apart from the audio of the scene, the video is also available. Following the line of research of this team during the two previous editions, an ASC system using squeeze-excitation techniques using Gammatone audio spectrograms and merged with a well-known architecture for computer vision (VGG16), has been proposed. The temporal structure of the data is handled by a recurrent architecture, performing data fusion from both modules both at an early and late stages. The results show that merging both sources of information makes the system more accurate, achieving a 15 percentage point accuracy improvement with respect to the Challenge baseline.
%However, the study and proposal of systems that are able to merge multiple sources of information is a field yet to be explored in the Machine Learning field.

\section{Acknowledgements}\label{sec:acknow}

\par This work is partially supported by ERDF and the Spanish Ministry of Science,
Innovation and Universities under Grant RTI2018-097045-B-C21, as well as grants AICO/2020/154 and AEST/2020/012 from Generalitat Valenciana.

% -------------------------------------------------------------------------
% Either list references using the bibliography style file IEEEtran.bst
\bibliographystyle{IEEEtran}
\bibliography{refs}

\end{sloppy}
\end{document}